\documentclass[10pt,twocolumn]{IEEEtran}

\usepackage{nopageno}
\usepackage{amsmath}
\usepackage{amsthm}
\usepackage{amssymb}
\usepackage{graphicx}
\usepackage{rotating}
\usepackage{verbatim}
\usepackage{enumitem}
\usepackage{ wasysym }

\usepackage[table]{xcolor}

\allowdisplaybreaks

\usepackage{xcolor}
\usepackage{tikz}

\newsavebox{\tempbox}
\usepackage{booktabs}
\usepackage{multirow}
\usepackage{xcolor}
\usepackage{etoolbox}
\usepackage{balance}

\DeclareMathOperator*{\argmin}{arg\,min}

\newcommand{\msf}[1]{\mathsf{#1}}
\newcommand{\mbf}[1]{\boldsymbol{\mathrm{#1}}}
\newcommand{\mcf}[1]{\mathcal{#1}}
\newcommand{\idc}[1]{\mathfrak{1}\left({\displaystyle #1}\right) }

\DeclareMathAlphabet\mbcf{OMS}{cmsy}{b}{n}

\newtheorem{theorem}{Theorem}

\newtheorem{lemma}[theorem]{Lemma}
\newtheorem{secass}[theorem]{Context}

\newtheorem{opdef}[theorem]{Operational Definition}

\usepackage{algorithm,algorithmic}

\definecolor{darkcamo}{HTML}{333c33} 
\definecolor{mediumcamo}{HTML}{83847a} 
\definecolor{lightcamo}{HTML}{bfb8ab} 
\definecolor{AFCgold}{HTML}{ffcf00}
\definecolor{AFCorange}{HTML}{d18a00}
\definecolor{AFCsilver}{HTML}{a1a1a3}
\definecolor{AFCgray}{HTML}{3f3f3f}
\definecolor{armygold}{HTML}{ffd530}

\newtheorem{remark}[theorem]{Remark}
\usetikzlibrary{shapes}

\IEEEoverridecommandlockouts

\providetoggle{cutty}
\settoggle{cutty}{false}

\begin{document}

\title{Optimal Update Policy for the Monitoring of Distributed Sources}

\author{
    Eric~Graves, 
        Jake B.~Perazzone, 
            Kevin Chan 
    \thanks{E. Graves, J. Perazzone, and K. Chan are with DEVCOM Army Research Laboratory, Adelphi, MD USA.}
    }
 
\maketitle
\thispagestyle{empty}

\begin{abstract}
When making decisions in a network, it is important to have up-to-date knowledge of the current state of the system.
Obtaining this information, however, comes at a cost.
In this paper, we determine the optimal finite-time update policy for monitoring the binary states of remote sources with a reporting rate constraint.
We first prove an upper and lower bound of the minimal probability of error before solving the problem analytically.
The error probability is defined as the probability that the system performs differently than it would with
full system knowledge.
More specifically, an error occurs when the destination node incorrectly determines which top-K priority sources are in the ``free'' state.
We find that the optimal policy follows a specific ordered 3-stage update pattern. 
We then provide the optimal transition points for each stage for each source.
\end{abstract}

\section{Introduction and Outline}

In order to make timely decisions in a system, up-to-date knowledge of its many time-varying processes must be known.
While \emph{Age of Information} (AoI) is a convenient measure of the timeliness of this knowledge, it is not always directly meaningful in the context of the system.
When estimating the value of a remote time-varying source, AoI is defined as the time elapsed since the remote source last sent an update to the destination.
This metric's effect on system performance, however, is not straightforward since the utility of each piece of information on decision-making is not directly quantified.
Therefore, directly minimizing AoI may or may not lead to optimal outcomes.

To be clear, though, when there is only a single Markovian remote source and the update policy must be independent of the source's value, AoI is the metric par excellence. 
Indeed, as discussed by Cover~\cite[Theorem~4]{cover1994processes}, for all discrete-time Markov processes, the divergence between the distribution over the possible states and the stationary distribution approaches zero with time.
In other words, as more time passes since the last update of the source's state, the destination's knowledge of the source becomes no better than a random guess.
By applying this reasoning, it is clear that the predictive power decreases as AoI increases.
In this scenario, minimizing\footnote{Usually, this is minimizing the average or maximum AoI over the run time of the system.} AoI yields the maximum predictive power of that remote source, which should, in theory, benefit the system regardless of whether the enhanced predictive power is necessary.

Nevertheless, when considering systems where either the update policy can be a function of the source's value or there are multiple Markovian remote sources competing to update, the superiority of AoI\footnote{Here, the extension is to consider the sum of the various AoIs.} is no longer assured. 
For the former, even a lack of an update provides information, while for the latter, evenly sharing the update resource does not necessarily provide utility to the system as a whole.
That is, having a more accurate view of a particular remote source is only helpful if that improved accuracy improves the use of the overall system.
It is possible for a policy that minimizes the AoI of all remote sources to, in turn, be providing unneeded fidelity to some sources and withholding it from others.
This can be partially addressed by weighted averaging of AoI, but its effect on the system is not straightforward.

For a simple example, consider a virtuous CEO (destination) who needs to schedule meetings to be up-to-date on the various company projects (remote source) for the purpose of being able to facilitate actions like project transitions at appropriate times.
Since each meeting consumes resources, e.g., time on the CEO's busy calendar, the optimal meeting schedule would prioritize both projects nearing some milestone as well as projects that the CEO deems of high value. 
Choosing a meeting schedule based on minimizing the sum of AoI across all projects, however, would lead to an unfavorable outcome.
This is because meetings for higher priority projects would be delayed in favor of a more balanced schedule where lower value projects will be scheduled simply because they have not given updates recently.
Additionally, the AoI formulation does not consider how the importance of each project may evolve over time.
Another simple example is job scheduling in large scale computing where some nodes are more favorable than others for a given task.

Before presenting an operational measure surrogate for AoI and a motivating example, we first discuss some of the history of AoI literature and other related works. 
The first to study age of information problems in the modern context is \cite{kaul2012real}.
Their formulation's objective was to minimize the AoI in a system containing a single remote source and destination connected by a queue with random service times. 
In particular, the queue is used to approximate an open network where multiple users may be slowing down transit of a packet to the final destination.
For this model, they showed that sending updates as quick as possible was not optimal since it would backlog the queue and thus delay newer updates from reaching the destination.
Since then, the age of information literature has expanded considerably \cite{yates2021age}.

Of particular importance is the class of AoI formulations under which the remote source can choose whether or not to update as a function of the value of the source.
The first to explore this research direction is \cite{sun2017update} which showed that, similarly to the base AoI case, zero-wait policies (i.e., send an update when the queue is empty) are not necessarily optimal. 
Instead, they determined optimal policies actually take the form of sending updates only when the queue is empty \emph{and} the cost (generally a function of the difference between the source's current and previous-update states) of not sending the update goes above a threshold.
Here, the intuition is that if the source has not changed significantly since the prior update, sending an update prematurely could delay a later update triggered from a rapid change in the source state.

Along these lines, there are two important works to consider.
First, \cite{sun2019sampling} studied the remote estimation of a Wiener process when updates about the process must go through a queue. 
Second, \cite{tsai2021unifying} studied a model similar to \cite{sun2019sampling}, but with a stochastic queue delaying the acknowledgment from destination to source.
These works once again determine a minimum time to wait before transmitting through an empty queue, for both source and destination, as being optimal.

On the other hand, when modeling a closed network, the random-delay queue becomes less applicable.
Indeed, with all nodes in the network being included in the model, most delay sources can be accounted for.
As a result, it is more appropriate for the queue to be replaced with a multiple access channel where the remote sources share the update resource.
Generally, this takes the form of limiting the number of remote sources that can update in any given time slot.

The work in \cite{he2016optimizing} is the first to consider such networks, using the sum of AoI terms as the metric. 
In their work, they showed that index policies were near optimal.
Index policies, whose origins date back to Gittins \cite{gittins1979bandit} and Whittle~\cite{whittle1988restless,whittle90index}, are ones in which each remote source has an associated index function whose input is the AoI as well as a threshold.
Updates are triggered when the index function output exceeds the threshold.
Index policies were later employed in the wireless-sensor-update scheduling framework \cite{kadota2018scheduling} which considered unreliable channels; \cite{Chen2022} which used source entropy instead of AoI; and \cite{kriouile2021global} which proved tighter bounds on the optimality gaps for index policies.

In this paper, we consider a system where remote source nodes update a destination node of their current state.
The goal is to minimize the probability that the destination incorrectly determines which top-$K$ priority sources are in the ``free'' state.
This model could, for example, model the monitoring and allocation of workers in a distributed computation system.
In the next section, we fully specify the system model and problem.
Then, we make the problem more tractable by estimating the objective function with an upper and lower bound.
An optimal policy based on the approximation is proven with the most interesting results being that the optimal policy follows a specific ordered 3-stage update pattern. 
We conclude by providing the optimal transition points for each stage for each source to complete the optimal update policy.

\section{Notation and Model}

\subsection{Notation}\label{sec:notation}

Random variables, constants,\footnote{Values of distinction will be written in capital letters using sans font so that they can be easier to recognize.} and sets will be written with upper case, lower case, and script, respectively.
For example, $X$ may take on value $x\in \mcf{X}$. 
Vectors will be denoted with bold, and subscript will be used to denote the particular coordinate; for instance, an $n$-dimensional random variable would be written as $\mbf{X} = (X_1,X_2,\dots,X_{n})$.
Subsets as coordinates denote the vector comprised of those coordinates, that is $\mbf{X}_{\mcf{A}} = (X_{a_1},\dots,X_{a_{q}})$ for $\mcf{A} = {a_1,\dots,a_{q}}$.

In the course of the work, there will be a few functions that occur often enough that they deserve explicit mention.
First, $\Pr \left( \cdot \right)$ denotes the probability of the event input into the function. 
Extensions of the probability function include conditional probability, where the conditional is to the right of the divider.
When dealing with random variables, say $X$ and $Y$, we may instead use $\mathbb{P}_{X|Y}(x|y)$, or even $\mathbb{P}(x|y)$, in place of $\Pr \left( X=x \middle| Y=y \right)$.
The expected value is denoted as $\mathbb{E}[X]$.
We define the norm operator as the $\ell^1$ norm, such that
$$\| \mbf{x}\| = \sum_{i=1}^{n} |x_i|.  $$
Finally, $[q] = \{1,\dots,q\}$ for any integer $q$. 
When referencing a substring such as $\{j,\dots,q\}$, we will write this as $[q]\setminus[j-1]$.

\tikzstyle{rect} = [draw, anchor=center, rectangle]
\tikzstyle{cir} = [draw, anchor=center, circle, text centered]
\tikzstyle{dia} = [draw,anchor=center, diamond]
\tikzstyle{coord} = [draw,anchor=center,star,star points=10, text centered]
\begin{figure}
\begin{center}

\begin{tikzpicture}[thick,scale=.8, every node/.style={transform shape}]

\foreach \x/\y/\index in {0/0/0}
{
\node[coord,label=above:{Destination}] (chq\index) at (\x,\y) {};
}

\foreach \x/\y/\index in {
-4/-2/1,
-2/-2/2,
0/-2/3,
4/-2/N}
{
\node[cir,label=below:{Source \index}] (apc\index) at (\x,\y) {};
}

\node at (2,-2) {\dots} ;

\draw[<-]  (chq0) to (apc1);
\draw[<-]  (chq0) -- (apc2);
\draw[<-]  (chq0) -- (apc3);
\draw[<-]  (chq0) -- (apcN);

\end{tikzpicture}
\caption{Network Model} \label{fig:flat}
\end{center}

\end{figure}
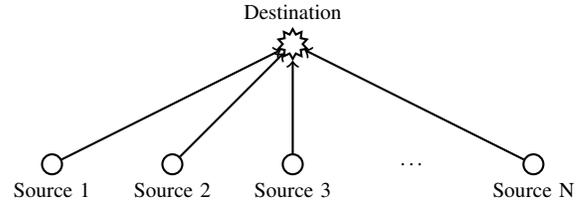

\subsection{Model}\label{sec:model}
 
We consider a flat network topology where one \emph{destination} node is directly connected with \emph{remote source} nodes $\mcf{N} = \{1,\dots,\msf{N}\}$, for a fixed integer $\msf{N}$, like that of Figure~\ref{fig:flat}. 

At times $t \in \{0,1,\dots,\msf{T}\}$, each remote source exists in either a ``busy'' state or a ``free'' state, denoted by the availability vector
$$ X_{n}(t) = \begin{cases} 
0 & \text{if node } n \text{ is busy} \\
1 & \text{if node } n \text{ is free},
\end{cases} $$
for each $n \in \mcf{N}$.
The state for each node evolves according to a Markov distribution with parameters $\mu$ and $\lambda$, in particular 
\begin{align} 
&\msf{P}_{X_n(t)|X_n(t-1)}(a|b) 
    \notag \\ & \quad 
    =
\begin{cases}
1-\lambda_n & \text{from free to free, i.e., } a = 1, b = 1 \\
 \lambda_n  & \text{from free to busy, i.e., } a = 0, b = 1 \\
\mu_n & \text{from busy to free, i.e., } a = 1, b = 0 \\
1-\mu_n & \text{from busy to busy, i.e., } a = 0, b = 0 .
\end{cases}
\end{align}
We assume $\mu_n,\lambda_n < .5$ for all $n\in \mcf{N}$, that is, the transitions occur due to some underlying continuous-time Markov distribution where the values for $\mu_n$ and $\lambda_n$ are the result of the time-discretization of this process. 
Also, we assume that each node starts at steady state, i.e., $\msf{P}_{X_n(0)}(1) = \frac{\mu_n}{\zeta_n}$, where $\zeta_n = \mu_n + \lambda_n$.

The remote source wishes to keep the destination informed about its status, and to do this, each remote source $n\in \mcf{N}$ may at any time\footnote{Note that this does not include $t=0$. This is done to allow for an initialization state. It is a relatively trivial matter and is only employed to make the results cleaner.} $t\in [\msf{T}] $ send an update to the destination that informs the destination of the value\footnote{This is for simplicity, considering the case where $X_n(t')$ for all $t' \leq t$ are sent back at time $t$ does not change results but makes for a more cumbersome presentation.} of $X_n(t)$. 
We assume that an update sent at time $t$ occurs after the state transition and is received by the destination at time $t+1$.
Note that in this model, the destination's understanding of the state is always delayed by at least $1$ time slot.
We denote whether or not remote source $n$ reports at time $t$ with the reporting vector
$$U_{n}(t) = \begin{cases} 
1 & \text{if source } n \text{ reports} \\
0 & \text{else}.
\end{cases} $$
Each remote source can use their full history to decide to send an update or not, thus $U_{n}(t)$ may be\footnote{As discussed in the next section, our goal is to design $U_n(t)$ and so it does not necessarily need to be dependent on any variables.} dependent on $\{X_{n}(i)\}_{i \in [t]\cup \{0\}}$ and $\{U_{n}(i)\}_{i \in [t-1]}$.
Finally, the monitoring vector $Y_n(t)$ denotes the most recent status the destination has of node $n$ at time $t$. 
Thus, we can write
$$ Y_n(t) = \begin{cases} X_n(t-1) & \text{if } U_n(t) = 1\\ Y_n(t-1) &\text{else}.\end{cases}$$
Note that $Y_n(t)$ is a deterministic function of $\{X_{n}(i)\}_{i \in [t]\cup \{0\}}$ and $\{U_{n}(i)\}_{i \in [t-1]}$ as this will be useful in the analysis.

\subsection{System Operation} 

The destination wishes to choose the $\msf{K}$ most desirable ``free'' remote sources at each time $t$. 
We assume for simplicity that the nodes are ordered by preference so that remote source node $1$ is more desirable than $2$ which is more desirable than $3$, and so on. 
In the event that less than $\msf{K}$ nodes are free, the destination instead wants to simply choose all free choices. 

As an example, for $\msf{N} = 6$, if the availability vector is $\mbf{X}(t) = [0,1,1,0,1,1]$, the top $\msf{K} = 3$ free choices would be $\{2,3,5\}$. 
In any case, to achieve this goal, the destination (rather na{\"i}vely\footnote{While choosing the set that is most likely free would improve the selector's performance, we have kept the destination's decision independent of this information.
In many real life settings, the design of the information processing system is independent of the communication protocols for getting that information.
That is the case here.}) will choose the top $\msf{K}$ free choices according to the monitoring vector $\mbf{Y}(t)$.
Each time, $t\in [\msf{T}]$, that the top $\msf{K}$ free remote sources differ according to $\mbf{X}(t)$ and $\mbf{Y}(t)$, they are said to be in error. 
Thus, continuing the earlier example, $\mbf{Y}(t) = [0,1,1,0,0,1]$ yields an error at time $t$, while $\mbf{Y}(t) = [0,1,1,0,1,0]$ would not, despite neither being equal to the earlier $\mbf{X}(t)$.
Unused nodes do not affect error.

With that in mind, our goal is to construct an update policy that minimizes the probability of error subject to an upper bound on the average number of updates that can be sent over the network. 
Each policy is then equivalent to specifying the values of $U_{n}(t)$ and their relationship. 
The operational measures for a policy are captured more formally in the following definitions. 
\begin{opdef}
The \emph{update rate} is
$$  \frac{1}{\msf{T}} \sum_{t=1}^{\msf{T}} \sum_{n = 1}^{\msf{N}}  \mathbb{E}[ U_n(t)] .$$
\end{opdef} 
\begin{opdef}
The \emph{time-averaged error probability} (to be referred to as \emph{error probability}) is
$$ \frac{1}{\msf{T}} \sum_{t=1}^{\msf{T}} \Pr \left( \mbf{X}_{[V(t)]}(t)  \neq \mbf{Y}_{[V(t)]}(t)  \right)   $$
where $V(t) = \min \left\{ n \in [\msf{N}] \, \middle| \, \|\mbf{X}_{[n]}(t)\| = \msf{K} \text{, else } n = \msf{N} \right\} $. 
\end{opdef}

Note, the error probability is equal to the expected ratio of times the system will perform differently than it would with full system knowledge. 
In other words, it is the probability at any given time that the system is operating incorrectly. 

Within the context of the systems operation, we want to determine the policy that minimizes the error probability for a given update rate. 
More specifically, for a given update policy $U_n$ and update rate constraint $r$, we wish to solve
\begin{align}\label{eqn:optProb}
    \min_{U_n (t)} \quad & \frac{1}{\msf{T}} \sum_{t=1}^{\msf{T}} \Pr \left( \mbf{X}_{[V(t)]}(t)  \neq \mbf{Y}_{[V(t)]}(t)  \right) \\
    \text{s.t.} \quad & \frac{1}{\msf{T}} \sum_{t=1}^{\msf{T}} \sum_{n = 1}^{\msf{N}}  \mathbb{E}[ U_n(t)] \leq r \, . \nonumber
\end{align}
Clearly, at one end, not updating will give the maximum error, while updating every change of status (requiring a rate of $\sum_{n=1}^{\msf{N}} \frac{2\mu_n\lambda_n}{\zeta_n}$) will give the minimum. 
One of our primary concerns is how the optimal error probability changes as a function of the update rate. 
This will, of course, require us to determine near optimal policies, which is the subject of the next section.

\section{Main Theorems}\label{sec:MT}

To characterize the update rate/error probability trade-off, we first approximate the error probability in a form that is analytically tractable.
Ideally, we would want to represent the system as a restless multi-arm bandit so that we could then solve it with well-known methods like Whittle's index. 
This is not possible, though, as a particular remote source's contribution to the error probability depends on the uncertainty around the other remote sources. 
That is, if a higher-priority remote source is guaranteed to be in error, then whether or not a lower-priority node is in error does not effect the error probability.

Still, when the error terms are small, the union bound allows for at least a partial representation of the error probability as a sum of the errors contributed by different remote sources. 
To that end, our first theorem provides upper and lower bounds on the error probability for a given $t$.
\newcounter{orderterm}
\setcounter{orderterm}{\thetheorem}
\begin{theorem}\label{thm:order_term}
\begin{equation}\label{eqn:errorBound}
\rho(t) \geq \Pr\left( \mbf{X}_{[V(t)]}(t) \neq \mbf{Y}_{[V(t)]}(t)  \right) \geq \frac{1}{4} \rho(t)
\end{equation}
where
\begin{align}
\rho(t) &= \min_{m \in [\msf{N}]} \rho(t,m):= \alpha_m + \sum_{n=1}^{m-1} \alpha_n \beta_n(t), \\
\text{and } \alpha_n &:= \Pr \left( \|X_{[n-1]}(t)\| < \msf{K}\right) = \Pr \left( \|X_{[n-1]}(0)\| < \msf{K}\right),
    \notag \\ 
\beta_n(t) &:= \Pr \left( X_n(t) \neq Y_n(t) \right).
    \notag
\end{align}
\end{theorem}
\begin{IEEEproof}
The full proof can be found in the Appendix~\ref{app:orderterm}. 

In essence, though, the theorem relies on representing the event $\mbf{X}_{[V(t)]}(t) \neq \mbf{Y}_{[V(t)]}(t)$ as a union of disjoint events, specifically $ \cup_{n=1}^{m} \mcf{E}_n(t)$, where 
$$\mcf{E}_n =\begin{array}{l} \{X_{[n-1]}(t)= Y_{[n-1]}(t)\} \\ \quad \cap \{\|X_{[n-1]}(t)\| < \msf{K} \} \\ \quad \cap \{ X_n(t) \neq  Y_n(t) \}. \end{array}$$
From there, the upper bound can be easily established using the union bound, while the lower bound requires first showing that
\begin{equation}
\Pr \left( \mcf{E}_n(t) \right) \geq \beta_n(t) \left( \alpha_n - \sum_{j=1}^{n-1} \alpha_j \beta_j(t) \right)^+ ,   
\end{equation}
followed by using optimization techniques to obtain the lower bound in \eqref{eqn:errorBound}. 
\end{IEEEproof}

\begin{remark}
For the remainder of the presentation we will use $\alpha_n$ and $\beta_n(t)$ as defined in Theorem \ref{thm:order_term}.
Furthermore, all sources $n \geq \argmin_m \rho(t,m)$ are referred to as being in the \emph{tail error}. 
This is done to make it easier to reference when a source is contributing to the error probability approximation. 
\end{remark}

Theorem~\ref{thm:order_term} allows for us to bound the minimum error probability by minimizing $\frac{1}{\msf{T}} \sum_{t=1}^{\msf{T}} \rho(t)$ instead.
Any policy that is good at reducing $\frac{1}{\msf{T}} \sum_{t=1}^{\msf{T}} \rho(t)$ should also be reasonably good at reducing the error probability.
It is worth noting that the contribution of a given remote source to the sum (either $\alpha \beta(t)$, $\alpha$, or $0$) offers an intuitive decomposition.
Specifically, when remote source $n$'s contribution is $\alpha_n\beta_n(t)$, i.e., nodes $n<m$, the $\alpha_n$ term can be viewed as the probability, at steady state, that remote source $n$'s information will change the system's response (in this case being included in the top $\msf{K}$) while the $\beta_n(t)$ term is the probability that the destination will be in error about remote source $n$'s status.
In a loose sense, then, the contribution of remote source $n$ is equal to the prior that remote source $n$ could cause an error multiplied by the probability that remote source $n$ is in error. 
On the other hand, for remote sources whose contribution is $\alpha$ or $0$, i.e., nodes $n\geq m$, can be viewed as the remote sources whose error is not well approximated by the union bound. 

While Theorem~\ref{thm:order_term} gives us a nice start to finding optimal solutions, it only considers a single instance of time, and thus depends on actions in the previous time step. 
This is probably best understood through an example, provided in Figure~\ref{fig:init}.
By applying Theorem~\ref{thm:order_term} at every time $t \in [\msf{T}]$, we are not guaranteed any particular ordering or structure to where the final term in the approximating summation will be.
In other words, the point at which sources below a certain priority will not contribute to an error event is not consistent.
Compare this to Figure~\ref{fig:simple} which is introduced later.

\begin{figure}
    \begin{center}
        \begin{tikzpicture}[thick,scale=.8, every node/.style={transform shape}]

\def\minx{1.5cm}
\def\miny{.75cm}
    
\node[rectangle,draw,minimum width=\minx,minimum height = \miny,anchor=west] at (\minx*0,0) {$\alpha_1 \beta_1(1)$};
\node[rectangle,draw,minimum width=\minx,minimum height = \miny,anchor=west] at (\minx*1,0) {$\alpha_1 \beta_1(2)$};
\node[rectangle,draw,minimum width=\minx,minimum height = \miny,anchor=west] at (\minx*2,0) {$\alpha_1 \beta_1(3)$};
\node[rectangle,draw,minimum width=\minx,minimum height = \miny,anchor=west] at (\minx*3,0) {$\alpha_1 \beta_1(4)$};
\node[rectangle,draw,minimum width=\minx,minimum height = \miny,anchor=west] at (\minx*4,0) {$\alpha_1 \beta_1(5)$};
\node[rectangle,draw,minimum width=\minx,minimum height = \miny,anchor=west] at (\minx*5,0) {$\alpha_1 \beta_1(6)$};

\node[rectangle,draw,minimum width=\minx,minimum height = \miny,anchor=west] at (\minx*0,\miny*-1) {$\alpha_2 \beta_2(1)$};
\node[rectangle,draw,fill = blue!20!white,minimum width=\minx,minimum height = \miny,anchor=west] at (\minx*1,\miny*-1) {$\alpha_2$};
\node[rectangle,draw,minimum width=\minx,minimum height = \miny,anchor=west] at (\minx*2,\miny*-1) {$\alpha_2 \beta_2(3)$};
\node[rectangle,draw,minimum width=\minx,minimum height = \miny,anchor=west] at (\minx*3,\miny*-1) {$\alpha_2 \beta_2(4)$};
\node[rectangle,draw,minimum width=\minx,minimum height = \miny,anchor=west] at (\minx*4,\miny*-1) {$\alpha_2 \beta_2(5)$};
\node[rectangle,draw,minimum width=\minx,minimum height = \miny,anchor=west] at (\minx*5,\miny*-1) {$\alpha_2 \beta_2(6)$};

\node[rectangle,draw,minimum width=\minx,minimum height = \miny,anchor=west] at (\minx*0,\miny*-2) {$\alpha_3 \beta_3(1)$};
\node[rectangle,draw,fill=gray!20!white,minimum width=\minx,minimum height = \miny,anchor=west] at (\minx*1,\miny*-2) {};
\node[rectangle,draw,minimum width=\minx,minimum height = \miny,anchor=west] at (\minx*2,\miny*-2) {$\alpha_3 \beta_3(3)$};
\node[rectangle,draw,fill = blue!20!white,minimum width=\minx,minimum height = \miny,anchor=west] at (\minx*3,\miny*-2) {$\alpha_3 $};
\node[rectangle,draw,minimum width=\minx,minimum height = \miny,anchor=west] at (\minx*4,\miny*-2) {$\alpha_3 \beta_3(5)$};
\node[rectangle,draw,fill = blue!20!white,minimum width=\minx,minimum height = \miny,anchor=west] at (\minx*5,\miny*-2) {$\alpha_3 $};

\node[rectangle,draw,minimum width=\minx,minimum height = \miny,anchor=west] at (\minx*0,\miny*-3) {$\alpha_4 \beta_4(1)$};
\node[rectangle,draw,fill=gray!20!white, minimum width=\minx,minimum height = \miny,anchor=west] at (\minx*1,\miny*-3) {};
\node[rectangle,draw,fill = blue!20!white,minimum width=\minx,minimum height = \miny,anchor=west] at (\minx*2,\miny*-3) {$\alpha_4$};
\node[rectangle,draw,fill=gray!20!white,minimum width=\minx,minimum height = \miny,anchor=west] at (\minx*3,\miny*-3) {};
\node[rectangle,draw,minimum width=\minx,minimum height = \miny,anchor=west] at (\minx*4,\miny*-3) {$\alpha_4 \beta_4(5)$};
\node[rectangle,draw,fill = gray!20!white,minimum width=\minx,minimum height = \miny,anchor=west] at (\minx*5,\miny*-3) {};

\end{tikzpicture}
    \end{center}
    \caption{An example of the summation terms approximating the error probability, with rows being unique to each remote source and columns representing time.}
    \label{fig:init}
\end{figure}

This affects the ability to directly apply something like Whittle's index.
Indeed, Whittle's index relies on the independence of the error contributions, but, here, decreasing the penalty from a given remote source at time $t$ may change the terms involved in the approximation of $\rho(t)$, thus creating dependence that is hard to account for.
Of course, one may assume that the remote source terms involved in calculating $\rho(t)$ are fixed and try to account for all variations, but that becomes an extremely difficult and challenging problem.

Instead, our next step in this work is to determine the form that optimal policies may take.
Doing so provides some much needed structure to any future optimization.
The next theorem simplifies the structure to a more analytical form; this structure will be discussed more after the theorem.
\begin{theorem} \label{thm:simp_form}
For each remote source $n\in \mcf{N}$, let $U_n$ be the policy that achieves 
$$\min \frac{1}{\msf{T}}\sum_{t=1}^{\msf{T}} \rho(t)$$
subject to $\frac{1}{\msf{T}} \sum_{n=1}^{\msf{N}} \sum_{t=1}^{\msf{T}} \mathbb{E}[U_n(t)] \leq r$ for a given $r>0$.  

For some integers $\msf{T}_n$, $n\in[\msf{N}]$, the policy, 
\small\begin{align}
\tilde  U_n(t) &= \begin{cases} 1 & t \in [\msf{T}_n] \,\text{and}\, X_n(t),Y_n(t) \in \{01,10\} \\
1 & t \in [\msf{T}]\backslash [\msf{T}_n] \,\text{and}\, X_n(t),Y_n(t) \in \{01\} \,\text{and}\, \lambda_n \geq \mu_n \\
1 & t \in [\msf{T}]\backslash [\msf{T}_n] \,\text{and}\, X_n(t),Y_n(t) \in \{10\} \,\text{and}\, \mu_n > \lambda_n \\
0 & \text{else},
\end{cases}\notag
\end{align}\normalsize
is such that 
\begin{equation}\notag
\frac{1}{\msf{T}} \sum_{t=1}^{\msf{T}} \tilde \rho(t)  \leq \frac{1}{\msf{T}}\sum_{t=1}^{\msf{T}} \rho(t) + \msf{O}\left( \frac{\msf{N}}{\msf{T}}\right) 
\end{equation}
for $\tilde \rho(t) = \min_{m\in [\msf{N}]} \alpha_{m} + \sum_{n=1}^{m-1} \alpha_n \Pr \left(  X_n(t) \neq \tilde Y_n(t) \right)$ with $\tilde Y_n(t)$ as the $Y_n(t)$ obtained when using policy $\tilde U$, and
\begin{equation}\notag 
\frac{1}{\msf{T}} \sum_{n=1}^{\msf{N}} \sum_{t=1}^{\msf{T}} \mathbb{E}[\tilde U_n(t)] \leq r + \msf{O}\left( \frac{\msf{N}}{\msf{T}} \right) .
\end{equation}

\end{theorem}
\begin{IEEEproof}
This theorem consists of a collection of multiple smaller lemmas, proved in Appendix~\ref{app:simp_form}.

In particular, Lemma~\ref{lem:simp_form:recur} shows that, for a fixed tail error function $f:[\msf{T}]\rightarrow [\msf{N}]$, the minimization of $\frac{1}{\msf{T}} \sum_{t=1}^{\msf{T}} \rho(t,f)$ subject to a given update rate can be written as a dynamic programming problem.  
From this, we learn directly that the optimal policy for each remote source will be dependent only upon the time $t$ and the availability and monitored values at that time, i.e., $X_n(t), Y_n(t)$.

We use this recursive formulation to analyze the structure of this solution. 
Lemma~\ref{lem:Delta_majorstay} says that if it is optimal for a remote source to update in a particular monitoring error state,\footnote{Not to be confused with a system error state, $\mbf{X}_{[V(t)]}(t)  \neq \mbf{Y}_{[V(t)]}(t)$, which affects the probability of error for the whole system.} e.g., $X_n(t),Y_n(t) = \{10\}$, at time $t+1$, then it is also optimal to update in that same error state at time $t$, as long as the remote source is not part of the error tail at time $t$.
Lemma~\ref{lem:dom} shows that, for each remote source, there exists an error state such that if it is optimal to update in this state, then it is also optimal to update in the reverse state; specifically, if $\lambda_n \geq \mu_n$, then updating in $X_n(t),Y_n(t)=10$ implies updating when $X_nY_n(t)=01$, and vice versa. 
Lemma~\ref{lem:dead} demonstrates that it is never optimal to update at time $t$ if the remote source is part of the error tail in time $t+1$. 

After establishing these lemmas, we optimize over the tail error. 
More specifically, for each source, we fix the number of time periods where a particular remote source is part of the tail error, and minimize the contribution of that remote source to the approximated error probability. 
With this approach, we show in Lemma~\ref{lem:ustayup} that for remote source $n$, this single-source-optimal policy has a parameter $\msf{T}_n$, where for times $t\in [\msf{T}_n]$, the optimal policy is to update in either error state, while for $t\notin [\msf{T}_n]$, the optimal policy updates in at most one of the error states. 
Given the policy dominance earlier, this implies that at most $1$ update occurs for $t >  \msf{T}_n$. 

Putting all of this together, it is easy to see that the policy stated in the lemma should be near optimal.
This is proved in Lemma~\ref{lem:error_calcs}.

\end{IEEEproof}

\begin{figure}
    \begin{center}
        \begin{tikzpicture}[thick,scale=.8, every node/.style={transform shape}]

\def\minx{1.5cm}
\def\miny{.75cm}

\node[rectangle,fill = orange!20!white,draw,minimum width=\minx,minimum height = \miny,anchor=west] at (\minx*0,0) {$\alpha_1 \beta_1(1)$};
\node[rectangle,fill = orange!20!white,draw,minimum width=\minx,minimum height = \miny,anchor=west] at (\minx*1,0) {$\alpha_1 \beta_1(2)$};
\node[rectangle,fill = orange!20!white,draw,minimum width=\minx,minimum height = \miny,anchor=west] at (\minx*2,0) {$\alpha_1 \beta_1(3)$};
\node[rectangle,fill = orange!20!white,draw,minimum width=\minx,minimum height = \miny,anchor=west] at (\minx*3,0) {$\alpha_1 \beta_1(4)$};
\node[rectangle,fill = orange!20!white,draw,minimum width=\minx,minimum height = \miny,anchor=west] at (\minx*4,0) {$\alpha_1 \beta_1(5)$};
\node[rectangle,fill = orange!20!white,draw,minimum width=\minx,minimum height = \miny,anchor=west] at (\minx*5,0) {$\alpha_1 \beta_1(6)$};

\node[rectangle,fill = orange!20!white,draw,minimum width=\minx,minimum height = \miny,anchor=west] at (\minx*0,\miny*-1) {$\alpha_2 \beta_2(1)$};
\node[rectangle,fill = orange!20!white,draw,minimum width=\minx,minimum height = \miny,anchor=west] at (\minx*1,\miny*-1) {$\alpha_2 \beta_2(2)$};
\node[rectangle,fill = orange!20!white,draw,minimum width=\minx,minimum height = \miny,anchor=west] at (\minx*2,\miny*-1) {$\alpha_2 \beta_2(3)$};
\node[rectangle,fill = orange!20!white,draw,minimum width=\minx,minimum height = \miny,anchor=west] at (\minx*3,\miny*-1) {$\alpha_2 \beta_2(4)$};
\node[rectangle,draw,minimum width=\minx,minimum height = \miny,anchor=west] at (\minx*4,\miny*-1) {$\alpha_2 \beta_2(5)$};
\node[rectangle,draw,fill = blue!20!white,minimum width=\minx,minimum height = \miny,anchor=west] at (\minx*5,\miny*-1) {$\alpha_2$};

\node[rectangle,fill = orange!20!white,draw,minimum width=\minx,minimum height = \miny,anchor=west] at (\minx*0,\miny*-2) {$\alpha_3 \beta_3(1)$};
\node[rectangle,draw,minimum width=\minx,minimum height = \miny,anchor=west] at (\minx*1,\miny*-2) {$\alpha_3 \beta_3(2)$};
\node[rectangle,draw,minimum width=\minx,minimum height = \miny,anchor=west] at (\minx*2,\miny*-2) {$\alpha_3 \beta_3(3)$};
\node[rectangle,draw,fill = blue!20!white,minimum width=\minx,minimum height = \miny,anchor=west] at (\minx*3,\miny*-2) {$\alpha_3 $};
\node[rectangle,draw,fill = blue!20!white,minimum width=\minx,minimum height = \miny,anchor=west] at (\minx*4,\miny*-2) {$\alpha_3 $};
\node[rectangle,draw,fill=gray!20!white,minimum width=\minx,minimum height = \miny,anchor=west] at (\minx*5,\miny*-2) {};

\node[rectangle,fill = orange!20!white,draw,minimum width=\minx,minimum height = \miny,anchor=west] at (\minx*0,\miny*-3) {$\alpha_4 \beta_4(1)$};
\node[rectangle,draw, minimum width=\minx,minimum height = \miny,anchor=west] at (\minx*1,\miny*-3) {$\alpha_4 \beta_4(2)$};
\node[rectangle,draw,fill = blue!20!white,minimum width=\minx,minimum height = \miny,anchor=west] at (\minx*2,\miny*-3) {$\alpha_4$};
\node[rectangle,draw,fill=gray!20!white,minimum width=\minx,minimum height = \miny,anchor=west] at (\minx*3,\miny*-3) {};
\node[rectangle,draw,fill=gray!20!white,minimum width=\minx,minimum height = \miny,anchor=west] at (\minx*4,\miny*-3) {};
\node[rectangle,draw,fill = gray!20!white,minimum width=\minx,minimum height = \miny,anchor=west] at (\minx*5,\miny*-3) {};

\end{tikzpicture}   
    \end{center}
    \caption{An example of the summation terms approximating the error probability after Theorem~\ref{thm:simp_form}. Here, orange boxes are ones where the source updates in either error state, while all other boxes the remote source only ever updates in one particular error state, determined by the values of $\mu_n$ and $\lambda_n$.}
    \label{fig:simple}
\end{figure}

Theorem~\ref{thm:simp_form} provides us a simpler form to optimize for. 
Instead of trying to optimize over sums that took the form of Figure~\ref{fig:init}, now we only need to consider optimizing over sums like that in Figure~\ref{fig:simple}.
The ordered sums are advantageous because each source only requires two parameters to determine its contribution to the approximated error probability; in particular, the number of time slots that the remote source updates in either error state (orange boxes), and the number of time slots that do not contribute to the approximated error probability (gray box).  
Most importantly, though, Theorem~\ref{thm:simp_form} and Theorem~\ref{thm:order_term} have exhausted the limits of what we can do by treating the contribution to the error probability of each remote source as independent.  

For our final theorem, we refine Theorem~\ref{thm:simp_form} and produce explicit $\msf{T}_n$ for each remote source that, in turn, determine the update policy. 
It should be noted that any policy obtained from this linear optimization problem would result in an error that is constant per source.
To see this, recognize that $\beta_n(t)$ should converge to its steady state according to an exponential function, the difference of which is a geometric series whose sum is constant. 
Furthermore, the error due to discretization of time would introduce a penalty of at most $2$ accounting for both places the discretization would impact.

\begin{theorem}\label{thm:oppcalc} 

For each remote source $n\in [\msf{N}]$, a sufficient $\msf{T}_n$ guaranteed by Theorem~\ref{thm:simp_form} is 
$$\msf{T}_n = \begin{cases} 
\msf{T} &  n \in \mcf{A}(\theta)   \\
\lfloor \msf{T}' \rfloor & n \in \mcf{B}(\theta) \setminus \mcf{A}(\theta) \\
0 &\text{else},
\end{cases}$$
where
\small\begin{align}
\mcf{A} (\theta) &= \left\{ n \in [\msf{N}] \middle| \theta < \alpha_n\left( \frac{1}{2\omega_n}-1 \right) \right\} \cap [\min \mcf{\tilde N}(\theta)-1],
    \notag \\ 
\mcf{B} (\theta) &= \left\{ n \in [\msf{N}] \middle| \theta \leq \alpha_n\left( \frac{1}{2\omega_n}-1 \right) \right\} \cap [\max \mcf{\tilde N}(\theta)-1],
    \notag \\ 
\mcf{\tilde N}(\theta) &=  \begin{cases} \argmin_{m\in \msf{N}} \tau(\theta,m) &  \min_{m\in [\msf{N}]} \tau(\theta,m) \leq 0 \\ \msf{N}+1 & \text{else},    \end{cases}
    \notag \\
\tau(\theta,m) &= \alpha_n - \sum_{n=m}^{\msf{N}} \frac{\nu_n}{\zeta_n} \min(\alpha_n,(\alpha_n+\theta)2\omega_n),
    \notag \\
\msf{T}' &= \msf{T} \frac{r -\sum_{n\in \mcf{A}(\theta)} \frac{2\mu_n \lambda_n}{\zeta_n}}{\sum_{n\in \mcf{B}(\theta)\setminus \mcf{A}(\theta)} \frac{2\mu_n \lambda_n}{\zeta_n}},
    \notag 
\\
\nu_n &= \min(\lambda_n,\mu_n),  \notag \\
\omega_n &= \max(\lambda_n,\mu_n),
    \notag 
\end{align}
\normalsize and where $\theta$ is the maximum real number  such that 
\small $$\sum_{n\in \mcf{A}(\theta)} \frac{2\mu_n \lambda_n}{\zeta_n} \leq r \leq \sum_{n\in \mcf{B}(\theta)} \frac{2\mu_n \lambda_n}{\zeta_n}    .$$ \normalsize
\end{theorem}
\begin{IEEEproof}
See Appendix~\ref{app:oppcalc} for the full proof.

Starting with the approximated policy of Theorem~\ref{thm:simp_form}, relaxing the assumption that the update time must be discrete, and replacing $\beta_n(t)$ with its steady state value $\frac{\nu_n}{\lambda_n}$ for $t > \msf{T}_n$ for each remote source $n$ (both of these relaxations incurring a penalty of at most $\frac{1}{\msf{T}}$), we can calculate the optimal values of $\msf{T}_n$ by solving the following optimization problem
\begin{equation} \label{eq:main:final_theorem}
\begin{array}{rll}
\min_{\mbf{s},\mbf{z} \in [0,\msf{T}]^{[\msf{N}]} \times [0,\msf{T}]^[\msf{N}]} &\sum_{n=1}^\msf{N} \epsilon_n(\mbf{s},\mbf{z})& \\
\text{s.t.} & s_n  \geq 0  &   \forall n \in [\msf{N}] \\
& z_n  \geq 0  & \forall n \in [\msf{N}] \\
&  s_{n} + z_{n} \leq s_{n-1} + z_{n-1}  & \forall n \in [\msf{N}] \\
& \sum_{n\in \mcf{N}} \frac{2\mu_n \lambda_n}{\zeta_n} s_n = r, & 
\end{array}
\end{equation}
where
$$\epsilon_{n}(\mbf{s},\mbf{z}) := \begin{array}{l}\alpha_n \frac{\nu_n}{\zeta_n} \left( 2 \omega_n s_n + z_n \right)  
\\ \quad + \alpha_n \left( s_{n-1} + z_{n-1} - s_{n} - z_{n} \right), \end{array}$$
and $s_{0} + z_0 = \msf{T}$ by definition. 
The above formulation is obtained after identifying $s_n$ as $\msf{T}_n$ and $ t> s_n + z_n$ as the region where $\tilde \alpha^\star_n(t) = 0$. 
Thus, $\sum_{n=1}^{\msf{N}}\alpha_n \left( s_{n-1} + z_{n-1} - s_{n} - z_{n} \right)$ represents the tail error introduced by Theorem~\ref{thm:order_term}.

The reported $\msf{T}_n$ are the values obtained solving Equation~\eqref{eq:main:final_theorem} using the Karush–Kuhn–Tucker (KKT) conditions.
Since the objective function and all constraints are linear, this solution is both necessary and sufficient. 
\end{IEEEproof}

\appendices 

\section{Proof of Theorem~\ref{thm:order_term}} \label{app:orderterm} 

Theorem~\ref{thm:order_term}'s proof relies first on representing the misselection event as a union of a number of mutually exclusive events and then approximating these events as independent. 
The approximation portion of the proof is not edifying but it is analytically significant.
To that end, before proving Theorem~\ref{thm:order_term}, it will be helpful to characterize 
\begin{equation}\label{eq:order_term:wfic3}
\begin{array}{lrl} 
f_k(w)= &\displaystyle \max_{ \{\beta_{i}\}_{i\in [k]} } &\displaystyle   \sum_{i=2}^{k} \sum_{ j=1 }^{i-1}  \alpha_j \beta_j \beta_i  
\\
&\text{s.t.} &\displaystyle \sum_{i=1}^k \alpha_i \beta_i = w \\
&&\displaystyle  \sum_{j=1}^{i-1} \alpha_j \beta_j \in  [0,\alpha_i] \quad \forall i \in [k]  \\
&&\displaystyle  \beta_i \in [0,1]  \quad \forall i \in [k]  
\end{array} 
\end{equation}
for $\{\alpha_{i}\in [0,1]\}_{i\in [k]}$. 
This characterization will be instrumental in comparing the misselection probability's lower bound ($f_k(w)$) with an upper bound ($w$). 

\begin{lemma} \label{lem:order_term_calc}
Define $f_k(w)$ as in Equation~\eqref{eq:order_term:wfic3} and let $c_{i}$ be defined recursively by 
\begin{equation} 
c_i = 4 \alpha_{i} \left( 1 - \frac{\alpha_i}{c_{i-1}} \right),\notag
\end{equation}
with $c_2 = 4 \alpha_2$.

For $w \in [0,\alpha_k]$ 
\begin{equation} 
f_k(w) = \frac{w^2}{c_{k}} \leq \frac{w}{2}.\notag 
\end{equation} 
\end{lemma}
\begin{IEEEproof}
This proof has been removed for space reasons. 
To prove this result, write 
\small\begin{align} 
f_i(w) &= \max_{ v \in [0,w] } \left( \frac{w - v}{\alpha_i} \right) x  + f_{i-1}(v) \label{eq:order_term:wficrecur}
\end{align}\normalsize
for $i\in [k]\setminus[1]$, and solve recursively. 

\end{IEEEproof}


\subsection{Proof of Theorem~\ref{thm:order_term}}
\begin{IEEEproof}

First observe that the misselection event at time $t$, denoted $\mcf{E}(t)$, can be represented as the union of disjoint events $ \cup_{i=1}^{n} \mcf{E}_i(t)$ where 
$$\mcf{E}_i =\begin{array}{l} \{X_{[i-1]}(t)= Y_{[i-1]}(t)\} \\ \quad \cap \{\|X_{[i-1]}(t)\| < k \} \\ \quad \cap \{ X_i(t) \neq  Y_i(t) \}. \end{array}$$
That is, there is a misselection if the estimated state of node $1$ is incorrect, or if the estimated state of node $1$ is correct but node $2$ is required and its state is incorrect, or if the estimated states of node $1$ and $2$ are correct but node $3$ is required and its state is incorrect, and so on. 
Since these events are disjoint, 
\begin{equation}\label{eq:order_term:ub_p0}
\Pr \left( \mcf{E}(t) \right) = \sum_{i=1}^{n} \Pr \left( \mcf{E}_i(t) \right) . 
\end{equation} 

To prove the upper bound
\begin{equation} \label{eq:order_term:ub}
\Pr \left( \mcf{E}(t) \right) \leq \min_{i \in [n]} \alpha_i + \sum_{j=1}^{i-1} \alpha_j \beta_j(t).
\end{equation}
first we establish for each $i \in [n]$ 
\begin{equation}\label{eq:order_term:ub_p1}
\Pr \left( \mcf{E}_i(t) \right) \leq \alpha_i \beta_i(t)
\end{equation} 
where $$\alpha_i(t) = \Pr \left( \|X_{[i-1]}(t)\| < k\right) = \Pr \left( \|X_{[i-1]}(0)\| < k\right)$$ and $\beta_i(t) = \Pr \left( X_i(t) \neq Y_i(t) \right)$.
This is somewhat trivial but follows directly as shown
\small\begin{align}
&\Pr \left( \mcf{E}_i(t) \right) 
    \notag \\ &\quad 
    = 
\Pr \left(    \begin{array}{l} \{X_{[i-1]}(t)= Y_{[i-1]}(t)\} \\ \quad \cap \{\|X_{[i-1]}(t)\| < k \} \\ \quad \cap \{ X_i(t) \neq  Y_i(t) \} \end{array} \right)  
    \label{eq:order_term:ub1} \\ & \quad
    =
\Pr \left(X_{i}(t) \neq Y_{i}(t) \right) \Pr \left(    \begin{array}{l} \{X_{[i-1]}(t)= Y_{[i-1]}(t)\} \\ \quad \cap \{\|X_{[i-1]}(t)\| < k \}  \end{array} \right)
    \label{eq:order_term:ub2} \\ & \quad 
    \leq
\Pr \left(X_{i}(t) \neq Y_{i}(t) \right) \Pr \left(    \cap \{\|X_{[i-1]}(t)\| < k \}  \right) 
    \label{eq:order_term:ub3}  \\ & \quad 
    =
\alpha_i \beta_i(t);
\end{align}\normalsize
where~\eqref{eq:order_term:ub2} is because $\{X_i,Y_i\}$ and $\{X_{[i-1]},Y_{[i-1]}\}$ are independent; and~\eqref{eq:order_term:ub3} follows because $\Pr \left( \mcf{A},\mcf{B} \right) \leq \min \left( \Pr \left(\mcf{A} \right) , \Pr \left( \mcf{B} \right) \right)$. 
At the same time, for any $i\in [n]$
\begin{equation}\label{eq:order_term:ub_p2}
\sum_{j=i}^n \Pr \left(  \mcf{E}_i(t) \right) \leq \Pr \left( \|X_{[i-1]}(t)\| < k \right) = \alpha_i.
\end{equation}
Combining Equations~\eqref{eq:order_term:ub_p0},~\eqref{eq:order_term:ub_p1}, and~\eqref{eq:order_term:ub_p2} yields~\eqref{eq:order_term:ub}.

For the lower bound observe that for each $i \in [n]$ 
\small\begin{align}
&\Pr\left( \mcf{E}_i(t) \right) 
    \notag \\ & \quad 
    =
\beta_i(t)  \Pr \left( X_{[i-1]}(t)= Y_{[i-1]}(t)  , \|X_{[i-1]}(t)\| < k  \right)  
     \\ & \quad 
    =
\beta_i(t) \left( \alpha_i - \Pr \left( \begin{array}{l} X_{[i-1]}(t)\neq Y_{[i-1]}(t)  , \\ \quad \quad \quad  \|X_{[i-1]}(t)\| < k \end{array} \right) \right)^+
    \label{eq:order_term:lb_a1}\\ & \quad 
    \geq
\beta_i(t) \left( \alpha_i - \sum_{j=1}^{i-1} \Pr \left( \mcf{E}_j(t) \right) \right)^+
    \label{eq:order_term:lb_a2}\\ & \quad 
    \geq
\beta_i(t) \left( \alpha_i - \sum_{j=1}^{i-1} \alpha_j \beta_j(t) \right)^+;
    \label{eq:order_term:lb_a3}
\end{align}\normalsize  
where~\eqref{eq:order_term:lb_a1} is by the law of total probability;~\eqref{eq:order_term:lb_a2} is because 
$$\{ X_{[i-1]}(t)\neq Y_{[i-1]}(t), \|X_{[i-1]}(t)\| < k \} \subset \cup_{j=1}^{i-1} \mcf{E}_i(t)$$
and $\{\mcf{E}_j(t)\}_{j \in [n]}$ are mutually exclusive; and finally~\eqref{eq:order_term:lb_a3} is by~\eqref{eq:order_term:ub_p1}. 
Hence, letting $i^\star$ be the minimum index such that $\alpha_{i^\star} \leq  \sum_{j=1}^{i^\star-1} \alpha_j \beta_j(t)$, from this observation we can obtain
\small\begin{align}
\Pr\left( \mcf{E}_i(t) \right) & \geq   \sum_{i=1}^n  \beta_i(t)  \left( \alpha_i - \sum_{j=1}^{i-1} \alpha_j \beta_j(t) \right)^+
     \\ &
    \geq 
\frac{1}{2} \sum_{i=1}^{i^\star-1} \alpha_i \beta_i(t) 
    \label{eq:order_term:lb_b1}
    \\ &
    \geq 
\frac{1}{4} \left( \alpha_{i^\star} + \sum_{i=1}^{i^\star-1} \alpha_i \beta_i(t) \right) 
    \label{eq:order_term:lb_b2}
    \\ &
    \geq 
\min_{i \in [n]} \frac{1}{4} \left( \alpha_{i} + \sum_{j=1}^{i-1} \alpha_j \beta_j(t) \right) ;
    \label{eq:order_term:lb_b3}
\end{align}\normalsize
where~\eqref{eq:order_term:lb_b1} is by Lemma~\ref{lem:order_term_calc}; and~\eqref{eq:order_term:lb_b2} is because $\alpha_{i^\star} \leq  \sum_{j=1}^{i^\star-1} \alpha_j \beta_j(t)$.

\end{IEEEproof}
\section{Proof of Theorem~\ref{thm:simp_form}}\label{app:simp_form}

\begin{lemma}\label{lem:simp_form:recur}

For each positive real number $r$, there exists a $\gamma$ such that the minimum of $\sum_{t = 1}^{\msf{T}} \rho(t,f(t))$, for a given function $f: [\msf{T}]\rightarrow [\msf{N}]$, subject to an update rate of $r$ is equal to
\begin{equation}
\sum_{a=0}^{1} \sum_{b=0}^{1} \tau^{(ab)}_n(1) \mathbb{P}_{X(1),Y(1)}(a,b),
\end{equation}
where the $\tau$ terms are derived recursively from the following equations 
\small\begin{align}
\alpha^\star_n(t) & = \alpha_n  \idc{n < f(t)} 
    \\ 
\tau^{(00)}_n(t-1) &= (1-\mu_n) \tau^{(00)}_{n}(t) + \mu_n \tau^{(10)}(t) 
    \\
\tau^{(01)}_n(t-1) &= \alpha_n^{\star}(t-1) + (1-\mu_n) \tau^{(00)}_{n}(t) + \mu_n \tau^{(10)}(t)
    \notag \\ & \quad 
+ \min \left( \gamma, \Delta^{(01)}_n(t) - \mu_n \Delta_n(t)  \right) 
    \\
\tau^{(10)}_n(t-1) &=  \alpha_n^{\star}(t-1) + (1-\lambda_n) \tau^{(11)}_{n}(t) + \lambda_n \tau_n^{(01)}(t) 
    \notag \\ & \quad 
+ \min \left( \gamma, \Delta^{(10)}_n(t) - \lambda_n \Delta_n(t) \right)
    \\ 
\tau^{(11)}_n(t-1) &= (1-\lambda_n) \tau^{(11)}_{n}(t) + \lambda_n \tau_n^{(01)}(t) 
    \\
\Delta_n(t)&=  \Delta^{(01)}_n(t) + \Delta^{(10)}_n(t)
    \\
\Delta^{(01)}_n(t) &= \tau^{(01)}_n(t) - \tau^{(00)}_n(t) 
    \\
\Delta^{(10)}_n(t) &= \tau^{(10)}_n(t) - \tau^{(11)}_n(t) 
\end{align}\normalsize
with the optimal policy being
\small\begin{align}
U_n^{(00)}(t) &= 0 \\
U_n^{(01)}(t) &= \idc{ \gamma < \Delta^{(01)}_n(t+1) - \mu_n \Delta_n(t+1) } \\
U_n^{(10)}(t) &= \idc{ \gamma < \Delta^{(10)}_n(t+1) - \lambda_n \Delta_n(t+1) } \\
U_n^{(11)}(t) &= 0 ,
\end{align}\normalsize
where $U^{(ab)}(t)$ represents $U(t)|\{XY(t-1) = ab\}$.    
\end{lemma}

\begin{IEEEproof}
This (somewhat trivial) proof has been removed for space reasons, only a sketch is provided

First, we use the Lagrange multiplier method to account for the update rate requirement, yielding
\begin{equation}
\min \sum_{t=1}^{\msf{T}} \alpha^\star(t) \beta(t) + \gamma \mathbb{E}[U(t)]
\end{equation}
where $\alpha^\star(t)$ is $\alpha$ if the remote source is not part of the tail error from Theorem~\ref{thm:order_term}. 
Writing the summand recursively as a dynamic program, noting 
\small\begin{align}
&\mathbb{P}_{XY(t)|XY(t-1)}(\{01\}\cup\{10\}|ab) 
    \notag \\
&\quad = \mathbb{P}_{X(t)|X(t-1)}(1-b|a) \mathbb{P}_{U(t-1)|XY(t-1)}(0|ab) 
    \notag \\ & \quad \quad \quad
+ \mathbb{P}_{X(t)|X(t-1)}(1-a|a) \mathbb{P}_{U(t-1)|XY(t-1)}(1|ab) 
\end{align}\normalsize
and similarly
\small\begin{align}
&\mathbb{E}[U(t-1|XY(t-1)=ab] = \mathbb{P}_{U(t-1)|XY(t-1)}(1|ab),
\end{align}\normalsize
yields the desired result.

\end{IEEEproof}

\begin{secass}\label{secass:0}
Fix $\gamma >0$ and function $f(t)$, let $U$, $\Delta$, and $\alpha^\star(t): \mcf{T}\rightarrow \{0,\alpha\}$ be the optimal values derived from Lemma~\ref{lem:simp_form:recur} for a particular remote source.

By $U^{(ab)}(t)$ denote the values of $U(t)$ given $XY(t) = ab$ for all pairs of $a,b\in\{0,1\}$. 
\end{secass}

\begin{lemma}\label{lem:dom}
$$\begin{cases} U^{(01)}(t) \geq U^{(10)}(t) & \lambda \geq \mu \\
U^{(10)}(t) \geq U^{(01)}(t)  & \mu > \lambda \end{cases}$$
for all $t \in [\msf{T}]$. 
\end{lemma}
\begin{IEEEproof}

Assume $\lambda \geq \mu$ with the alternative case being symmetrical. 

The lemma follows by proving for all $t \in [\msf{T}]$
\begin{equation}\label{eq:dom:1}
\omega(t) \geq \upsilon(t) \quad \text{and} \quad \omega(t) \geq 0 
\end{equation}
where
\begin{align*}
\omega(t) &= \Delta^{(01)}(t) - \mu \Delta(t)  \\
\upsilon(t) &= \Delta^{(10)}(t) - \lambda \Delta(t)  .
\end{align*}
Clearly Lemma~\ref{lem:dom} is a direct consequence of Equation~\eqref{eq:dom:1} since $ U^{(01)}(t) = \idc{\gamma< \omega(t+1)}$ and $U^{(10)}(t) = \idc{\gamma < \upsilon(t+1)}$.

This can be achieved via mathematical induction. 
In particular for the base case
\small\begin{align} 
\omega(\msf{T}) - \upsilon(\msf{T}) &= (1-2\mu)\alpha^\star(\msf{T}) - (1-2\lambda) \alpha^\star(\msf{T}) 
    \\ & 
    = 
2(\lambda - \mu) \alpha^\star(\msf{T}) \geq 0 ,
\end{align}\normalsize 
where the final step follows because $\lambda \geq \mu$ and $\alpha^\star(\msf{T}) \geq 0$;
and 
\small\begin{align} 
\omega(\msf{T}) &= (1-2\mu) \alpha^\star(\msf{T}) \geq 0 
\end{align}\normalsize
where the final line follows because $\mu \leq \frac{1}{2}$.

For the induction step assume that 
\small\begin{align} 
\omega(t+1) - \upsilon(t+1) \geq 0   \quad \text{and} \quad \omega(t+1) \geq 0.
\end{align}\normalsize
With these assumptions it follows that
\small\begin{align}
\omega(t) &= (1-2\mu) \alpha^\star(t) + (1-\mu)\min(\gamma,\omega(t+1)) 
    \notag \\ & \quad \quad 
- \mu \min(\gamma,\upsilon(t+1)) 
    \\ & 
    \geq 
 (1-2\mu) \min( \gamma, \omega(t+1)) \geq 0 
\end{align}\normalsize
since $\mu \leq \frac{1}{2}$ and $\min(\gamma,\upsilon(t+1)) \leq \min(\gamma,\omega(t+1))$.
Similarly 
\small\begin{align}
\omega(t) - \upsilon(t) &\geq 2( \lambda - \mu) \left[ \alpha^\star(t)  + \min(\gamma, \omega(t+1)) \right]
    \label{eq:dom:4} \\ &  
    \geq 0 
    \label{eq:dom:5}  ;
\end{align}\normalsize
where~\eqref{eq:dom:4} is because $1-\mu + \lambda \geq \frac{1}{2}$ and $\min(\gamma,\upsilon(t+1)) \leq \min(\gamma,\omega(t+1));$
and~\eqref{eq:dom:5} is because $\omega(t+1) \geq 0 $ and $\alpha^\star(t) \geq 0 $. 

Therefore~\eqref{eq:dom:1} follows by induction, and the lemma follows from~\eqref{eq:dom:1}.
\end{IEEEproof}

\begin{lemma}\label{lem:Delta}
For all $t \in [\msf{T}]$ 
$$\Delta(t)  \leq \frac{2 \alpha}{\zeta} .$$
\end{lemma}
\begin{IEEEproof}
We will use induction for the proof. 

To that end, note the result is trivial for $t = \msf{T}$.

So, assume that $\Delta(t+1) \leq \frac{2 \alpha}{\zeta}$, and observe that
\small\begin{align}
\Delta(t) &= 2 \alpha^\star(t) + \min( \gamma, \Delta^{(01)}(t+1) - \lambda \Delta(t+1))
    \notag \\ & \quad \quad \quad  
+ \min( \gamma, \Delta^{(10)}(t+1) - \mu \Delta(t+1))
    \\ &
    \leq 
2 \alpha + \Delta(t+1) (1- \lambda - \mu) 
    \\ & 
    \leq 
\frac{2 \alpha}{\zeta};
\end{align}\normalsize
where the final step uses the induction assumption. 
\end{IEEEproof}

\begin{lemma}\label{lem:Delta_majorstay}
Suppose $\alpha^\star(t+1) = \alpha$,  
\begin{equation} \notag
\text{if } U^{i}(t)(t+1) = \alpha + \gamma \text{ then } U^{i}(t) = 1
\end{equation}
for $i\in \{(01),(10)\}$. 
\end{lemma}
\begin{IEEEproof}
Suppose that $\lambda \geq \mu$ and therefore $\Delta^{(01)} \geq \left| \Delta^{(10)} \right|^+$ by Lemma~\ref{lem:dom}.

Now, $U^{(01)}(t+1) = 1$ implies that $\Delta^{(01)}(t+1) = \alpha^{\star}(t+1) + \gamma$, hence
\small\begin{align}
\Delta^{(01)}(t+1) - \lambda \Delta(t+1)  &\geq \alpha^{\star}(t+1) + \gamma - \frac{2 \lambda}{\zeta}\alpha 
    \label{eq:Delta_majorstay:1} \\ &
    >
\gamma  
    \label{eq:Delta_majorstay:2};
\end{align}\normalsize
where Equation~\eqref{eq:Delta_majorstay:1} is by Lemma~\ref{lem:Delta} and~\eqref{eq:Delta_majorstay:2} is because $\mu > \lambda$ and the assumption $\alpha^\star(t+1) = \alpha$.

Still assuming $\mu \geq \lambda$, note that if $U^{(10)}(t+1) = 1$, this implies that 
\begin{equation}
\frac{1- 2\mu}{2\mu} \alpha > \gamma
\end{equation}
since $U^{(10)}(t+1) = 1$ requires $(1-\mu) \Delta^{(10)}(t+2) - \mu \Delta^{(01)}(t+2) > \gamma$ hence
\small\begin{align}
\gamma &< (1-\mu) \Delta^{(10)}(t+2) - \mu \Delta^{(01)}(t+2) 
    \label{eq:Delta_majorstay3} \\ &
    \leq
(1-2\mu) \Delta^{(10)}(t+2) 
    \label{eq:Delta_majorstay4} \\ & 
    \leq
(1-2\mu)(\alpha+\gamma).
\end{align}\normalsize
Note now that, because $U^{(01)}(t+1)=1$ if $U^{(10)}(t+1) =1$, we have
\small\begin{align}
U^{(10)}(t) &= \idc{ \gamma < \Delta^{(10)}(t+1) - \mu \Delta(t+1)} 
    \\ &
    = 
\idc{ \gamma < (1-2\mu)(\alpha+\gamma)} = 1.
\end{align}\normalsize

A symmetrical proof follows for $\lambda \geq \mu$.
\end{IEEEproof}

\begin{lemma}\label{lem:dead}
For the optimal policy given fixed $\alpha$, if $\alpha^\star(t+1) = 0$ then $U^{(10)}(t) = U^{(01)}(t) = 0$.
\end{lemma}
\begin{IEEEproof}

Assume $\mu \geq \lambda$ with a symmetrical proof for the alternative case following.
In this case we only need to show $U^{(01)}(t) = 0$, since this already implies $U^{(10)}(t) = 0$.

To this end note 
\small\begin{align}
\Delta^{(01)}(t+1) - \lambda \Delta(t) \leq \Delta^{(01)}(t+1) \leq \alpha^{\star}(t+1) + \gamma = \gamma
\end{align}\normalsize
and hence
\begin{equation}
U^{(01)}(t) = \idc{ \gamma < \Delta^{(01)}(t+1) - \lambda \Delta(t)} = 0 .
\end{equation}
\end{IEEEproof}

\begin{secass}\label{secass:1}
For a fixed $\gamma >0$ and set of integers $\{\msf{A}_n\}_{n \in \mcf{N}}$, where $\msf{A}_1 \leq \msf{A}_2 \leq \dots \leq \msf{A}_n$, for each remote sources $n\in \mcf{N}$ let $U_n$ and $\alpha^\star_n(t): \mcf{T}\rightarrow \{0,\alpha_n\}$ be the policy that achieves
$$\min \sum_{n=1}^\mcf{N} \sum_{t=1}^{\msf{T}} \alpha_n^\star(t) \beta_n(t) + \gamma \mathbb{E}[U_n(t)],$$
where $\beta_n(t) = \Pr \left( X_n(t) = Y_n(t) \right)$ and $Y(t)$ is defined by policy $U_n(t)$, and the minimum is subject to
$\sum_{t=1}^{\msf{T}} (\alpha_n - \alpha^\star_n(t)) = \msf{A}_n$.

By $U_n^{(ab)}(t)$ denote the values of $U_n(t)$ given $X_nY_n(t) = ab$ for all pairs of $a,b\in\{0,1\}$.

For all of the following lemmas, we will only deal with a single remote source and thus drop the subscript to avoid clutter.
\end{secass}
\begin{lemma} \label{lem:ustayup}
Assume Context~\ref{secass:1}. 
If $U^{(01)}U^{(10)}(t) = 11 $ for some $t\in [\msf{T}]$, then there exists a $\msf{T}' \in [\msf{T}]$ such that $U^{(01)}U^{(10)}(t) = 11$ if and only if $t \in [\msf{T}'].$
\end{lemma}
\begin{IEEEproof}
This proof has been removed for space concerns, instead a proof sketch is provided. 

If the lemma were not true, then there would exist a time $t'$ such that $U^{(01)}U^{(10)}(t) = 11 $ and $U^{(01)}U^{(10)}(t-1) = 00$ then it must be that $\alpha^\star(t) = 0$, or else Lemma~\ref{lem:Delta_majorstay} would be violated. 
The alternative policy though
\begin{equation}
\tilde \alpha^\star(t) = 
\begin{cases} 
\alpha^\star(t) & t \in [\hat t -1] \\
\alpha^\star(t+1) & t\in [\msf{T}-1]\setminus [\hat t-1]\\ 
\alpha^\star(\hat t) & t = \msf{T}
\end{cases}
\end{equation}
and
\small\begin{align}
\tilde U^{(01)} \tilde U^{(10)}(t) = 
\begin{cases}
U^{(01)}U^{(10)}(t) & t \in [\hat t -2] \\
U^{(01)}U^{(10)}(t+1) & t\in [\msf{T}-1]\setminus [\hat t-2]\\ 
U^{(01)}U^{(10)}(\hat t-1) & t = \msf{T}
\end{cases}
\end{align}\normalsize
yield the same error probability as the original but with a smaller update rate, contradicting the assumption.

\end{IEEEproof}

\begin{lemma}\label{lem:error_calcs}

Define the alternative policy $\tilde U$ by $\tilde U^{(00)}(t) = \tilde U^{(11)}(t) = 0$ for all $t$ and
\small\begin{align}
\tilde U^{(01)} \tilde U^{(10)}(t) &= \begin{cases}
11 & t \in [\msf{T}'] \\
10 & t \in [\msf{T}]\setminus [\msf{T}'] \text{ and } \lambda \geq \mu \\
01 & t \in [\msf{T}]\setminus [\msf{T}'] \text{ and } \lambda  < \mu 
\end{cases} \notag
\end{align}\normalsize
where $\msf{T}'$ is the value guaranteed by Lemma~\ref{lem:ustayup}, and define function $\tilde \alpha^\star: [\mcf{T}] \rightarrow \{0,\alpha\} $ by
\small\begin{align} 
\tilde \alpha^\star(t) = \begin{cases} \alpha & t \in [\msf{T}-\msf{A}] \\
0 & t \in [\msf{T}]\setminus[\msf{T}-\msf{A}]
\end{cases} \notag.
\end{align}\normalsize 

Here,
\small\begin{align}
\sum_{t=1}^{\msf{T}} \tilde \alpha^\star(t) \tilde \beta(t) \leq O(1) + \sum_{t=1}^{\msf{T}}  \alpha^\star(t)  \beta(t),\notag \\
\sum_{t=1}^{\msf{T}} \mathbb{E}[ \tilde U(t)]  \leq 1 + \sum_{t=1}^{\msf{T}} \mathbb{E}[U(t)] .\notag 
\end{align} \normalsize
\end{lemma}
\begin{IEEEproof}

This proof has been removed due to space concerns only sketch has been provided. 

As in previous proofs assume $\lambda \geq \mu$, noting that the alternative's proof is symmetric. 

Essential to the proof is that the alternative policy forces the monitored value to go to $0$. 
Indeed, updates only occurs when $XY(t) = 01$, and afterwards $Y(t+1) = 0$ and therefore no more updates can occur. 
Thus, when comparing the optimal policy and our alternative, the alternative policy can only have $1$ more update than the optimal.

To prove the bound on the error term relies on the fact that the steady state for a policy with no updates and the steady state for the alternative policy are
\small \begin{equation}
\mathbb{P}^\star \left(  \begin{matrix} 0,0\\ 0,1\\ 1,0 \\ 1,1 \end{matrix} \right)  =  \left[ \begin{matrix} \frac{\lambda}{\zeta} \mathbb{P}_Y(0) \\ \frac{\lambda}{\zeta} \mathbb{P}_Y(1) \\ \frac{\mu}{\zeta} \mathbb{P}_Y(0) \\ \frac{\mu}{\zeta} \mathbb{P}_Y(1)  \end{matrix} \right] \quad \text{and} \quad \mathbb{\tilde P}^\star \left(  \begin{matrix} 0,0\\ 0,1\\ 1,0 \\ 1,1 \end{matrix} \right) = \left[ \begin{matrix} \frac{\lambda}{\zeta}  \\ 0 \\ \frac{\mu}{\zeta}  \\ 0  \end{matrix} \right]
\end{equation} \normalsize
respectively.

Thus, over larger values of $\msf{T}$ long sequences of not updating move the distribution in a direction with a larger error $(\mathbb{\tilde P}_{XY}(01) + \mathbb{\tilde P}_{XY}(10))$ than that of the alternative policy. 
Over large enough $n$, convergence of Markov chains guarantees our result. 

\end{IEEEproof}

\section{Theorem~\ref{thm:oppcalc}}\label{app:oppcalc}
\begin{IEEEproof}

To show this result, we need to solve the following optimization problem 
\begin{equation} \label{eq:final_max:opt_start}
\begin{array}{rll}
\min_{\mbf{s},\mbf{z}} &\sum_{i=1}^{\msf{N}} \epsilon_i(\mbf{s},\mbf{z})& \\
\text{s.t.} & s_i  \geq 0  &   \forall i \in [\msf{N}] \\
& z_i  \geq 0  & \forall i \in [\msf{N}] \\
&  s_{i} + z_{i} \leq s_{i-1} + z_{i-1}  & \forall i \in [\msf{N}] \\
& \sum_{i=1}^{\msf{N}} \frac{2\lambda_i \mu_i}{\zeta_i} s_i = \msf{T}r &
\end{array}
\end{equation}
where $ s_{0} + z_{0} = \msf{T}$ and
\begin{equation}
\epsilon_{i}(\mbf{s},\mbf{z}) := \alpha_i\frac{\nu_i}{\zeta_i} \left( 2 \omega_i s_i + z_i \right)  + \alpha_i \left( s_{i-1} + z_{i-1} - s_{i} - z_{i} \right) .
\end{equation}
The Kaush-Kuhn-Tucker (KKT) conditions can be used to solve Equation~\eqref{eq:final_max:opt_start} since both the function and the constraints are linear.

To that end, let 
\begin{equation}
f= \begin{array}{l} 
\sum_{i=1}^{\msf{N}} \epsilon_i(\mbf{s},\mbf{z})   + \sum_{i=1}^{\msf{N}} \chi_i (-s_i) + \sum_{i=1}^{\msf{N}} \psi_i (-z_i) \\
\quad  + \sum_{i=1}^{\msf{N}} \xi_i (s_{i} + z_{i} -  s_{i-1} - z_{i-1}) \\ 
\quad + \theta ( - \msf{T} r + \sum_{i=1}^{\msf{N}} \frac{2\lambda_i\mu_i}{\zeta_i} s_i) 
\end{array}.
\end{equation}
By the KKT conditions the optimal policy must satisfy the stationary conditions\footnote{These following from \small\begin{align}
\frac{\partial f}{\partial s_i} &= \alpha_i - (\alpha_{i} - \alpha_{i+1}) \kappa_{i} - \chi_{i} -  \xi_{i+1} \kappa_i + \xi_i \kappa_i + \theta  \notag \\
\frac{\partial f}{\partial z_i} &= -\alpha_{i} \frac{\omega_i}{\zeta_i} + \alpha_{i+1} - \psi_i + \xi_i - \xi_{i+1}. \notag 
\end{align}\normalsize}  
\small\begin{align}
\chi_{i}  + \xi_{i+1} - \xi_i - (\alpha_i + \theta) \frac{2\lambda_i \mu_i}{\zeta_i} + \alpha_i - \alpha_{i+1} &=0   \quad \forall i \in [\msf{N}] \label{eq:final_max:kkt1} \\ 
\psi_i + \xi_{i+1} - \xi_i   + \alpha_{i} \frac{\omega_i}{\zeta_i} - \alpha_{i+1} & = 0  \quad \forall i \in [\msf{N}] \label{eq:final_max:kkt2}
\end{align}\normalsize
where $\alpha_{\msf{N}+1} , \xi_{\msf{N}+1} = 0 $; the complementary slackness conditions 
\small\begin{align}
\chi_i(-s_i)  &= 0  \quad    \forall i \in [\msf{N}] 
    \label{eq:final_max:kkt:cs:chi}\\
\psi_i( -z_i)  &=0 \quad  \forall i \in [\msf{N}] 
    \label{eq:final_max:kkt:cs:psi} \\
\xi_i\left(  s_i+z_i  - s_{i-1} - z_{i-1} \right) &= 0 \quad   \forall i \in [\msf{N}]  
    \label{eq:final_max:kkt:cs:xi} ; 
\end{align}\normalsize
the primal feasibility conditions 
\small\begin{align} 
-s_i   &\leq 0  \quad    \forall i \in [\msf{N}] \label{eq:final_max:kkt:pf:x}\\
-z_i   &\leq 0  \quad    \forall i \in [\msf{N}] \label{eq:final_max:kkt:pf:y}\\
s_i + z_i - s_{i-1} - z_{i-1} & \leq 0  \quad \forall i \in [\msf{N}] \label{eq:final_max:kkt:pf:xy} \\
\sum_{i=1}^{\msf{N}} \frac{2 \lambda_i \mu_i}{\zeta_i}   s_i &= \msf{T} r   \label{eq:final_max:kkt:pf:xlr};
\end{align}\normalsize
and the dual feasibility conditions  
\small\begin{align}
\chi_i  \geq 0   \quad 
\psi_i  \geq 0 \quad  
\xi_i  \geq 0 \quad \forall i \in [\msf{N}] \label{eq:final_max:kkt:df}. %
\end{align}\normalsize 

Solving the KKT conditions will show that for a fixed $\theta$, the minimum occurs at 
\small\begin{align}
s_i &= \sigma_i t_i \quad  \forall i \in [\msf{N}]
    \label{eq:final_max:final_x} \\
z_i &= (1-\sigma_i) t_i \quad \forall i \in [\msf{N}]
    \label{eq:final_max:final_y} 
\end{align}\normalsize
for any values $\{\sigma_i,t_i\}_{i \in [\msf{N}]}$ that satisfy
\small\begin{align}
\sigma_i & \in \begin{cases}
\{0\}&  \theta > \alpha_i \left( \frac{1}{2\omega_i} - 1\right) \\
[0,1] & \theta =  \alpha_i \left( \frac{1}{2\omega_i} - 1\right) \\
\{1\} & \theta < \alpha_i \left( \frac{1}{2\omega_i} - 1\right),
\end{cases} 
    \label{eq:final_max:final_z}\\
t_i & \in \begin{cases}
\{t_{i-1}\} & i \in [\msf{N}]\setminus\mcf{I}^\star(\theta) \\
\{0\} & i = \iota_{+}(\theta) \\
[0,t_{i-1}] & i \in \mcf{I}^\star(\theta)\setminus \{\iota_{+}(\theta)\},
\end{cases}
    \label{eq:final_max:final_t}\\
\sum_{i=1}^{\msf{N}} \frac{2\lambda_i\mu_i}{\zeta_i} \sigma_i t_i &= \msf{T} r,
\end{align}\normalsize
where $t_0:=0$ and $\iota_{+}(\theta)  = \max \mcf{I}^\star(\theta) $ and\footnote{Recall that $\alpha_{\msf{N}+1} :=0$. The inclusion of this term in the sum is so that $\iota_{+}(\theta) = \msf{N}+1$ when all terms are not negative.} 
\begin{equation*}
\mcf{I}^\star(\theta) := \argmin_{i\in [\msf{N}+1]} \left| \alpha_{i} - \sum_{j=i}^{\msf{N}+1} \frac{\nu_j}{\zeta_j} \min \left( \alpha_j , (\alpha_j + \theta) 2\omega_j \right)\right|^-.
\end{equation*}

To show that~\eqref{eq:final_max:final_x} and~\eqref{eq:final_max:final_y} are indeed the case, it will be helpful to establish a few things.
First, observe that 
\begin{equation} \label{eq:final_max:ass1:0}
\xi_i = \begin{array}{l} \alpha_{i} - \sum_{j=i}^{\msf{N}} \frac{\nu_j}{\zeta_j} \min \left( \alpha_j , (\alpha_j + \theta) 2\omega_j \right) \\ \quad + \sum_{j=i}^{\msf{N}} \min\left(  \chi_j, \psi_j \right)  \end{array},
\end{equation}
by combining Equations~\eqref{eq:final_max:kkt1} and~\eqref{eq:final_max:kkt2}.
Next, for any $k \in \mcf{I}^{\star}(\theta)$ and index $i\in [\msf{N}+1]\setminus[k]$
\begin{equation} \label{eq:final_max:assassass}
\frac{\xi_{k} - \xi_i}{i-k}  \leq \max_{j\in [i-1]\setminus[k-1]} \min\left(  \chi_j, \psi_j \right) ,
\end{equation}
with the inequality being strict if $i \notin \mcf{I}^{\star}(\theta)$.
Equation~\eqref{eq:final_max:assassass} can be derived as follows
\small\begin{align}
\xi_{k} - \xi_i 
 &
    \leq
\sum_{j=k}^{i-1} \min\left(  \chi_j, \psi_j \right)
    \label{eq:final_max:ass1:3}\\ & 
    \leq 
(i-k) \max_{j \in [i-1]\setminus[k-1]} \min\left(  \chi_j, \psi_j \right) 
    \label{eq:final_max:ass1:4}
\end{align}\normalsize
where~\eqref{eq:final_max:ass1:3} follows from $k\in \mcf{I}^\star(\theta)$.
Note the inequality in~\eqref{eq:final_max:ass1:3} is strict if $i \notin \mcf{I}^{\star}(\theta)$. 
Also, for any $k \in \mcf{I}^{\star}(\theta)$ and $i \in [k-1]$,
\begin{equation}\label{eq:final_max:minxi}
\xi_{k} \leq \xi_{i}
\end{equation}
with the inequality being strict for $i\notin \mcf{I}^\star(\theta)$. 
Equation~\eqref{eq:final_max:minxi} follows similarly to~\eqref{eq:final_max:assassass}.
Finally,
\begin{equation}\label{eq:final_max:minchipsi}
 \min\left(  \chi_k, \psi_k \right) = 0 \quad  \forall k \in [\iota_{+}(\theta)-1].
\end{equation}
Indeed, assume $\min\left(  \chi_k, \psi_k \right)>0$ for some $k \in [\iota_{+}(\theta)-1]$ then
\begin{equation}
\xi_i = \xi_{\iota_{+}(\theta)} + \sum_{j=i}^{\iota_{+}(\theta)-1} \min\left(  \chi_j, \psi_j \right)  \geq \min\left(  \chi_k, \psi_k \right)>0
\end{equation}
for all $i\in [k]$
using Equation~\eqref{eq:final_max:ass1:0} and the dual feasibility conditions~\eqref{eq:final_max:kkt:df}.
Thus $\min\left(  \chi_k, \psi_k \right)>0$ for some $k \in [\iota_{+}(\theta)-1]$ would mean that $\xi_{i}>0$ for $j \in [i]$ which causes a contradiction.
In specific, noting $\min\left(  \chi_k, \psi_k \right)>0$ requires $s_k = z_k = 0$, 
$$s_k =z_k = s_{k-1} = z_{k-1} = \dots = s_1 = z_1 = 0  $$
by conditions~\eqref{eq:final_max:kkt:cs:xi},~\eqref{eq:final_max:kkt:cs:chi}, and~\eqref{eq:final_max:kkt:cs:psi} 
while simultaneously $s_1 + z_1 = t$ by condition~\eqref{eq:final_max:kkt:cs:xi} and $\xi_1 >0$.
Hence the assumption that $\min\left(  \chi_k, \psi_k \right)>0$ for some $k \in [\iota_{+}(\theta)-1]$ must be false.

We begin by proving that, except in the degenerate case\footnote{In such a case, $[\msf{N}]\setminus[\iota_{+}(\theta)-1]$ is empty.} where $\iota_{+}(\theta) = n+1$, 
\begin{equation}\label{eq:final_max:xy_gte}
s_{i} = z_{i} = 0  \quad \forall i \in [\msf{N}]\setminus[\iota_{+}(\theta)-1].
\end{equation}
Of course, this can also be written as $s_i = \sigma_i t_i$ and $z_i = (1-\sigma_i)t_i$ for $\sigma_i \in [0,1]$ and $t_i = 0$. 
Let $k \in [\msf{N}+1]\setminus[\iota_{+}(\theta)]$ be the smallest index such that\footnote{The existence of this index is guaranteed since $\xi_{n+1} = 0$ by definition.} $\xi_k = 0$. 
For this index $k \notin \mcf{I}^{\star}(\theta)$ since $k > \iota_{+}(\theta)$, hence 
\begin{equation}
0 \leq  \frac{\xi_{\iota_{+}(\theta)} - \xi_k}{k- \iota_{+}(\theta)} < \max_{j \in [k-1]\setminus[\iota_{+}(\theta)-1]} \min\left(  \chi_j, \psi_j \right)
\end{equation}
by Equation~\eqref{eq:final_max:assassass}.
Letting $j'\in [k-1]\setminus[\iota_{+}(\theta)-1]$ be the maximizing coordinate, we now have that $s_{j'}=z_{j'}=0$ by conditions~\eqref{eq:final_max:kkt:cs:chi} and~\eqref{eq:final_max:kkt:cs:psi}. 
Furthermore we know that $\xi_{j''} >0$ for all $j'' \in [j']\setminus[\iota_{+}(\theta)]$ since $k$ was the smallest index such that $\xi_k = 0$.
Equation~\eqref{eq:final_max:xy_gte} now follows since
\begin{equation} \label{eq:final_max:bigxy}
s_{j'} = z_{j'} = s_{j'-1} = z_{j'-1} = \dots = s_{\iota_{+}(\theta)} = z_{\iota_{+}(\theta)} = 0
\end{equation}
by condition~\eqref{eq:final_max:kkt:cs:xi}.

Before moving on to the case of $i < \iota_{+}(\theta)$, it will be helpful to prove that 
\begin{equation}\label{eq:final_max:xistar}
\xi_{\iota_{+}(\theta)} = 0,
\end{equation}
or by combining~\eqref{eq:final_max:ass1:0},~\eqref{eq:final_max:minchipsi}, and~\eqref{eq:final_max:xistar} the more applicable 
\begin{equation}
\xi_{i} = 0 \quad \forall i \in \mcf{I}^{\star}(\theta).
\end{equation}
Equation~\eqref{eq:final_max:xistar} can be seen as a consequence of~\eqref{eq:final_max:xy_gte} and~\eqref{eq:final_max:minxi}.
That is, if $\xi_{\iota_{+}(\theta)} >0$, then $\xi_{j} >0$ for all $j\in [\iota_{+}(\theta)]$.
This would simultaneously imply $s_1 + z_1 = 0$ by~\eqref{eq:final_max:xy_gte} and~\eqref{eq:final_max:kkt:cs:xi} due to $\xi_{j}>0$ for $j\in [\iota_{+}(\theta)]\setminus [1]$; as well as implying $s_1 +z_1 = \msf{T}$ by~\eqref{eq:final_max:kkt:cs:xi} due to $\xi_1 >0$.
Hence, it must be that $\xi_{\iota_{+}(\theta)} = 0$.

Now, moving on to prove Equations~\eqref{eq:final_max:final_x} and~\eqref{eq:final_max:final_y} for $i<\iota_{+}(\theta)$.
Here
\small\begin{align}
\xi_{j} &= \alpha_{j} - \alpha_{\iota_{+}(\theta)} - \sum_{i=j}^{\iota_{+}(\theta)-1} \frac{\nu_i}{\zeta_i} \min \left( \alpha_i , (\alpha_i + \theta) 2\omega_i \right)
    \label{eq:final_max:final_xi} \\
\chi_j &= \frac{2\omega_j \nu_j}{\zeta_j} \left| \theta - \alpha_j \left( \frac{1}{2\omega_j} - 1\right) \right|^+ 
    \label{eq:final_max:final_chi}\\
\psi_j &= -\frac{2\omega_j \nu_j}{\zeta_j} \left| \theta - \alpha_j \left( \frac{1}{2\omega_j} - 1\right) \right|^- ,
    \label{eq:final_max:final_psi}
\end{align}\normalsize
can be obtained by combining~\eqref{eq:final_max:ass1:0},~\eqref{eq:final_max:minchipsi}, and~\eqref{eq:final_max:xistar}.
Recalling that $\xi_{i}=0$ if and only if $i \in \mcf{I}^{\star}(\theta)$ it must follow that 
\small\begin{align}
 s_i + z_i &  \in \begin{cases}  
\{ s_{i-1} + z_{i-1}\} & i \notin \mcf{I}^{\star}(\theta) \\ 
[0, s_{i-1} + z_{i-1}] & i \in \mcf{I}^{\star}(\theta).
\end{cases}
\end{align}\normalsize
Furthermore, by using Equations~\eqref{eq:final_max:final_chi},~\eqref{eq:final_max:final_psi},~\eqref{eq:final_max:kkt:cs:chi}, and~\eqref{eq:final_max:kkt:cs:psi} it follows that for $i$ such that $\theta > \alpha_i \left( \frac{1}{2\omega_i} - 1\right)$
\small\begin{align}
s_i = 0  \quad z_i= s_i + z_i;
\end{align}\normalsize
while for $i$ such that $\theta > \alpha_i \left( \frac{1}{2\omega_i} - 1\right)$
\small\begin{align}
s_i = s_i+z_i  \quad  z_i = 0 ;
\end{align}\normalsize
and for the remaining case of $\theta = \alpha_i \left( \frac{1}{2\omega_i} - 1\right)$ we trivially have 
\small\begin{align}
s_i  \in [0,s_i+z_i]  \quad  z_i \in [0,s_i + z_i].
\end{align}\normalsize
Putting these equations together it follows that all $\{s_i,z_i\}_{i\in [\msf{N}]}$ that solve the KKT conditions are a subset of the set of values detailed in~\eqref{eq:final_max:final_x} and~\eqref{eq:final_max:final_y}.

That all values $\{s_i,z_i\}_{i\in [\msf{N}]}$ defined by Equations~\eqref{eq:final_max:final_x} and~\eqref{eq:final_max:final_y} are equal can be easily seen by plugging those value into the approximated and simplifying to
\small\begin{align}
\sum_{i=1}^{\msf{N}}\epsilon_{i}(\mbf{s},\mbf{z})   &= - \msf{T} \theta r + \msf{T} \alpha_{\iota_{-}(\theta)} 
    \notag \\ & \quad \quad \quad 
+  \msf{T} \sum_{i=1}^{\iota_{-}(\theta)-1} \frac{\nu_i}{\zeta_i} \min \left( \alpha_i, (\alpha_i + \theta) 2 \omega_i \right),
\end{align}\normalsize
where $\iota_{-}(\theta) = \min \mcf{I}^{\star}(\theta)$, by using that 
the various KKT conditions as well as 
\begin{equation}
\sigma_i \alpha_i 2 \omega_i + (1-\sigma_i) \alpha_i = \min\left(\alpha_i, (\alpha_i+\theta)2\omega_i \right) - \theta  2 \omega_i \sigma_i  
\end{equation}
for all $i\in [\msf{N}]$ and that 
\small\begin{align}
0 &= \xi_i - \xi_j \\
&= \alpha_{i} - \alpha_{j} - \sum_{k=i}^{j-1} \frac{\nu_k}{\zeta_k} \min \left( \alpha_k , (\alpha_k + \theta) 2\omega_k \right) \\
&=  - \sum_{k=i}^{j-1} \frac{\nu_k}{\zeta_k} \min \left( \alpha_k , (\alpha_k + \theta) 2\omega_k \right) + \alpha_{k+1} - \alpha_k 
\end{align}\normalsize
for all $i\in \mcf{I}^\star(\theta)\setminus[\iota_{+}(\theta)]$ and corresponding $j = \min \mcf{I}^{\star}(\theta)\setminus [i]$ (that is the smallest value in $\mcf{I}^{\star}(\theta)$ greater than $i$).



\end{IEEEproof}

\bibliographystyle{IEEEtran}
\bibliography{bib} 

\end{document}